\documentclass[apj]{emulateapj} 
\usepackage{mathptmx} 
\usepackage[colorlinks=true,urlcolor=blue,citecolor=blue,linkcolor=blue]{hyperref}
\newcommand{\be}{\begin{equation}}
\newcommand{\ee}{\end{equation}}
\def\kms{km\,s$^{-1}$}
\def\dd{\mathrm{d}}
\def\Halpha{\mbox{H\hspace{0.1ex}$\alpha$}}
\def\rmit#1{{\it #1}}
\def\eg{\rmit{e.g.,}}
\def\ie{\rmit{i.e.,}}
\def\edt#1{#1}
\def\rmd{{\rm d}}

\def\rme{{\rm e}}

\def\exp{\rme}

\def\is{\ensuremath{\!=\!}}

\def\eh2{\ensuremath{e_{\mathrm{H2}}}}
\def\nh2{\ensuremath{n_{\mathrm{H2}}}}
\def\exp{\rme}

\def\h2{\ensuremath{\mathrm{H}_2}}

\def\HeI{\ion{He}{1}}
\def\CaII{\ion{Ca}{2}}

\def\MgI{\ion{Mg}{1}}
\def\MgII{\ion{Mg}{2}}
\def\MgIIk{\ion{Mg}{2}\,k}
\def\MgIIh{\ion{Mg}{2}\,h}
\def\MgIII{\ion{Mg}{3}}
\def\hk{{h\&k}}
\def\MgIIhk{\MgII\, \hk}
\def\Bifrost{{\it Bifrost}}

\def\multitd{{\it Multi3d}}

\def\RH{{\it RH}}
\def\figspath{.}

\def\kthree{\mbox{k$_3$}}    
\def\hthree{\mbox{h$_3$}}
\def\ktwo{\mbox{k$_2$}}
\def\htwo{\mbox{h$_2$}}
\def\kone{\mbox{k$_1$}}     
\def\hone{\mbox{h$_1$}}

\def\ktwov{\mbox{k$_{2V}$}}
\def\ktwor{\mbox{k$_{2R}$}}

\def\htwov{\mbox{h$_{2V}$}}
\def\htwor{\mbox{h$_{2R}$}}

\def\Ktwov{\mbox{K$_{2v}$}}
\def\Htwov{\mbox{H$_{2v}$}}

\begin{document}

\title{The formation of IRIS diagnostics \\
II. The formation of the MG\ II\ H\&K lines in the solar
atmosphere}
  
   \author{J.~Leenaarts$^{1}$}\email{jorritl@astro.uio.no}
   \author{T. M. D. Pereira$^{2,3,1}$}\email{tiago.pereira@astro.uio.no}
   \author{M.Carlsson$^{1}$}\email{mats.carlsson@astro.uio.no}
   \author{H. Uitenbroek$^{5}$}\email{huitenbroek@nso.edu}
   \author{B. De Pontieu$^{3,1}$}\email{bdp@lmsal.com}

\affil{$^1$ Institute of
  Theoretical Astrophysics, University of Oslo, P.O. Box 1029
  Blindern, N--0315 Oslo, Norway}
\affil{$^2$NASA Ames Research Center, Moffett Field, CA 94035, USA}
\affil{$^3$Lockheed Martin Solar \& Astrophysics Lab,
         Org.\ A021S, Bldg.\ 252, 3251 Hanover Street
         Palo Alto, CA~94304 USA}
\affil{$^4$NSO/Sacramento Peak P.O. Box 62
         Sunspot, NM 88349--0062 USA}

   \date{Received; accepted}

   \begin{abstract}
NASA's Interface Region Imaging Spectrograph (IRIS) small explorer mission will
study how the solar atmosphere is energized. IRIS contains an imaging
spectrograph that covers the \MgIIhk\ lines as well as a slit-jaw
imager centered at \MgIIk. Understanding the observations requires
forward modeling of \MgIIhk\ line formation from 3D radiation-MHD
models. This paper is the second in a series where we undertake this
modeling. We compute the vertically emergent \hk\ intensity
from a snapshot of a dynamic 3D radiation-MHD model of the solar
atmosphere, and investigate which diagnostic information about the
atmosphere is contained in the synthetic line profiles.

We find that the Doppler shift of the central line depression
correlates strongly with the vertical velocity at optical depth unity,
which is typically located less than 200\,km below the transition
region (TR). By combining the Doppler shifts of the h and the k line we
can retrieve the sign of the velocity gradient just below the TR.  The
intensity in the central line depression is anticorrelated with the
formation height, especially in subfields of a few square Mm. This
intensity could thus be used to measure the spatial variation of the
height of the transition region.

The intensity in the line-core emission peaks correlates with the
temperature at its formation height, especially for strong emission
peaks. The peaks can thus be exploited as a temperature
diagnostic. The wavelength difference between the blue and red peaks
provides a diagnostic of the velocity gradients in the upper
chromosphere. The intensity ratio of the blue and red peaks correlates
strongly with the average velocity in the upper chromosphere.

We conclude that the \MgIIhk\ lines are excellent probes of the very
upper chromosphere just below the transition region, a height regime
that is impossible to probe with other spectral lines. They also
provide decent temperature and velocity diagnostics of the middle
chromosphere.

\end{abstract}

  
\section{Introduction}                          \label{sec:introduction}

This is the second paper in the series exploring the diagnostic
potential of the \MgIIhk\ lines in preparation for NASA's
Interface Region Imaging Spectrograph (IRIS) space mission.
In paper I 
%
%
we discuss the feasibility of modeling
these lines in non-LTE and with partial frequency redistribution (PRD)
in increasingly realistic Radiation-Magneto Hydrodynamic (RMHD)
simulations. We find that the h\&k resonance lines can be modeled
reliably with a 4-level plus continuum atomic model of \MgII, comprised
of the $3s\,^2\!S$ ground state, the two $3p\,^2\!P$ levels, an
artificial level representing higher energy levels to correctly reproduce
the \MgII--\MgIII\ ionization balance, and the \MgIII\ ground state.
The calculations presented in paper~I also make clear that PRD always
needs to be accounted for. For wavelengths in the line wings up to and
including the \ktwo\ and \htwo\ peaks, 1D column-by-column radiative
transfer suffices to accurately reproduce profile shapes and their
spatial variations. For line-core wavelengths between the
emission peaks 3D transfer is required, but complete redistribution
(CRD) is sufficient.

Because of its large abundance and the dominance of its singly ionized
state the \MgIIhk\ lines sample higher layers of the solar chromosphere
than other common diagnostics of that region, namely \Halpha, the 
\CaII\,H\&K lines, the \CaII\ IR triplet lines and the \HeI\ 1083 nm
subordinate line.

\begin{figure}
  \includegraphics[width=8.8cm]{\figspath/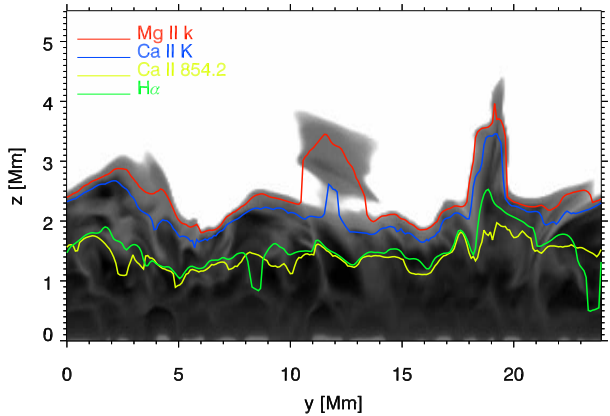}
  \caption{Differences between $\tau \is 1$ heights of \MgIIk,
    \CaII\,K, \CaII\,854.2\,nm and \Halpha\ in an $yz$-slice of the 3D
    model atmosphere. The image displays the temperature, clipped at
    20\,kK, with curves of the maximum $\tau \is 1$ height of the
    various lines overplotted.
    \label{fig:diag_comp}}
\end{figure}

The difference in formation height of several of these diagnostics as expressed 
by the maximum height where $\tau \is 1$ over the line profile at each
location and calculated through a $yz$-slice from the 3D RMHD atmosphere
employed in Paper I is plotted in Figure~\ref{fig:diag_comp}.
The CaII\,K, \CaII\,854.2\,nm (as representative of the IR triplet)
and \Halpha\ formation heights
\citep[see][]{2012ApJ...749..136L}
were computed in 3D non-LTE using \multitd.
Clearly \MgIIk\ and \CaII\ K have very similar formation properties,
with the main difference between them the higher opacity due to the
larger solar abundance of magnesium
\citep[7.60 for Mg and 6.34 for Ca,][]{2009ARA&A..47..481A},
which, all else equal, translates to a difference in $\tau \is 1$
height of 2.9 scale heights. The larger magnesium opacity leads also to a
larger thermalization depth and thus magnesium retains temperature
sensitivity in the source function to larger heights.
This sensitivity is reflected in the fact that \MgIIk\ shows central
reversals in the line core everywhere
\citep{2001ApJ...557..854M}
while 
\citet{2008A&A...484..503R} 
report that 25\% of the internetwork \CaII\,H profiles show a
reversal-free line core. 

Observations in all lines mentioned above show elongated fibrilar structure that most
likely are to a large extent aligned with the magnetic field. All
lines can thus be used to study morphology. Yet, observations in one
single line alone cannot yield a sufficiently complete picture of
physical processes in the chromosphere due to the limits in formation
height range and the usability of each line to infer all relevant
physical parameters: mass density, temperature, velocity and magnetic
field.

So far, multi-wavelength studies of the chromosphere
have mainly utilized \Halpha\ and  \CaII\,854.2\,nm
\citep[\eg,][]{2009A&A...503..577C,2012ApJ...752..108S,2009A&A...500.1239R},
which in our model atmosphere form roughly at
equal height, approximately the same height as optical depth unity for the
\MgIIhk\ emission peaks (see Figure~\ref{fig:h2_k2_plot_1}).

The two lines are, however, used very differently. 
The \Halpha\ line core intensity and width are density and temperature
diagnostics
\citep{2012ApJ...749..136L},
while the \CaII\,854.2\,nm line can be used as velocity and
temperature diagnostic, and, if observing the full Stokes vector at
sufficient sensitivity, magnetic field strength and orientation
\citep{2012A&A...543A..34D}.
Since, in the weak field limit
\citep[][p.~258]{stenflo1994solar},  
the amount of circular polarization due to the Zeeman effect
in a spectral line is proportional to the ratio of the line splitting
over the Doppler width, and the amount of linear polarization
proportional to the square of that ratio,
we can expect only very weak polarization signals in any of the other 
chromospheric lines because the h\&k and H\&K lines are too far to the
blue and H$\alpha$ has a Doppler width that is too large
\citep{2011ASPC..437..439U}.
The only other truly chromospheric spectral line that can be 
used effectively for magnetometry is the \HeI\ 1083 nm line,
which can be used to measure strength and orientation of weak
and strong magnetic fields because of its sensitivity to both the
Zeeman and Hanle effects
\citep{2008ApJ...683..542A}.

Through its capability of obtaining spectra of the \MgIIhk\ lines
at high spatial, spectral and temporal resolution the IRIS mission
will provide a unique opportunity
to extend multi-layer studies of the solar chromosphere to the
region just below the transition region, which cannot be probed
by other means. In this paper we focus on determining what 
diagnostic information can be obtained from observation of
\MgIIhk\ spectra by forward modeling of these lines through the same snapshot
of a 3D RMHD simulation used in
\citet{2012ApJ...749..136L}
and in paper I.
Among other findings we show that the magnesium 
resonance lines provide excellent
velocity diagnostics through the Doppler shift of \hthree\ and
\kthree\ and an indication of the spatial variation of the height of
the transition region through the intensity of \hthree\ and \kthree.

The structure of the paper is as follows: In
Section~\ref{sec:model_atmos}--\ref{sec:rad_trans} we describe our 3D
model atmosphere and the radiative transfer
computations. Section~\ref{sec:data_reduction} describes the reduction
of the synthetic data to the intensity and Doppler-shift of the
central emission peaks (\ktwo\ and \htwo) and the central minima
(\kthree\ and \hthree), the parameters that we use to characterize the
line profiles.  In Section~\ref{sec:hk3_diag}--\ref{sec:hk2_diag} we
describe how these parameters correlate with properties of the model
atmosphere. In Section~\ref{sec:indprof} we describe the formation of
three example profiles in detail to illustrate the correlations that
we found. We finish with a discussion of the results and our
conclusions in Section~\ref{sec:conclusions}.

\section{Model atmosphere} \label{sec:model_atmos}

We study the \hk\ line formation in a  snapshot of a 3D radiation-MHD
simulation performed with the \Bifrost\ code
\citep{2011A&A...531A.154G}.
The same snapshot has been used by
\citet{2012ApJ...749..136L}
to investigate \Halpha\ line formation and was also used in Paper~I. A 2D slice through this
atmosphere was used by
\citet{2012ApJ...758L..43S}
to investigate depolarization of scattered light by the Hanle effect
in the Lyman-$\alpha$ line.

\Bifrost\ solves the equations of resistive MHD on a staggered Cartesian
grid with a selection of modules describing various physical processes. The simulation
we use here included radiative transfer in four multi-group opacity bins
including coherent scattering affecting the energy balance in the
photosphere and low chromosphere
\citep{1982A&A...107....1N, 2000ApJ...536..465S, 2010A&A...517A..49H}
parametrized radiative losses in
the upper chromosphere, transition region and corona
\citep{2012A&A...539A..39C}, thermal
conduction along magnetic field lines \citep{2011A&A...531A.154G}
and an equation of state that
includes the effects of non-equilibrium ionization of hydrogen
\citep{2007A&A...473..625L}.

The simulation covers a physical extent of $24 \times 24 \times
16.8$\,Mm$^3$, with a grid of $504 \times 504 \times 496$ cells, extending
from 2.4\,Mm below the average height of $\tau_{500}=1$ to
14.4\,Mm above covering
the upper convection zone, photosphere, chromosphere and the lower
corona.  The horizontal axes have an equidistant grid spacing of 48\,km, the vertical grid
spacing is non-uniform, with a spacing of 19\,km between $z=-1$ and
$z=5$\,Mm. The spacing increases towards the bottom and top of the
computational domain to a maximum of 98\,km. The simulation contains a magnetic field with a
an average unsigned strength of 50\,G in the photosphere, concentrated
in the photosphere in two clusters of opposite polarity 8 Mm apart. For further
details on this snapshot we refer to 
\citet{2012ApJ...749..136L}.
%

\section{Radiative transfer computations} \label{sec:rad_trans}

We perform non-LTE radiative transfer computations using the
4-level-plus-continuum model atom from paper~I with two different
codes: The first is a version of \RH\ by
\citet{2001ApJ...557..389U},
modified to use MPI so it can efficiently solve the radiative
transfer in all columns in the 3D model atmosphere assuming each column
is a plane-parallel 1D atmosphere, including angle-dependent partial
redistribution.

The second code we use is \multitd\
\citep{2009ASPC..415...87L}.
This is an MPI-parallelized radiative transfer code that can
evaluate the radiation field in full 3D, taking the horizontal
structure in the model atmosphere into account. For
this 3D computation we reduced the grid size of the model atmosphere to
$252\times252\times200$ grid cells to reduce the amount of
computational work. 

The \MgIIhk\ lines are influenced by both PRD and 3D effects. We lack
the capability to compute both simultaneously, so for the analysis of
the line profiles we use the hybrid approach described and motivated
in Paper I: we use 1D PRD computations with \RH\ for the line profile
up to and including the central emission peaks. The intensity of the
central line depression is taken from the 3D CRD computation with
\multitd.

\section{Reduction of the synthetic spectra} \label{sec:data_reduction}

\subsection{Line parameters}
 
\begin{figure}
  \includegraphics[width=8.8cm]{\figspath/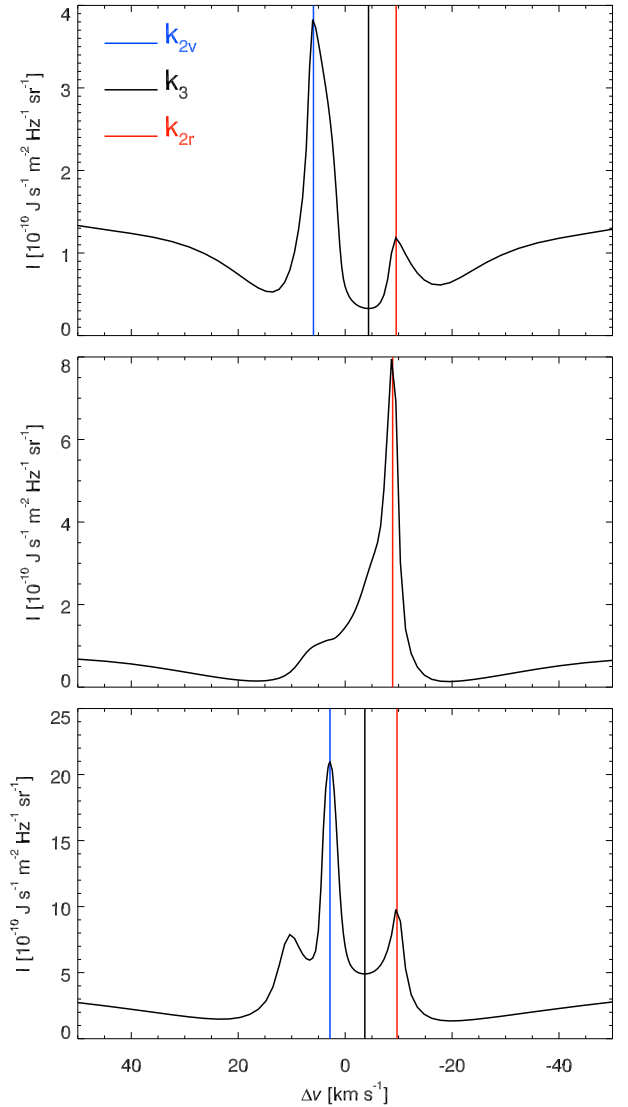}
  \caption{Example \MgIIk\ line profiles with the identification of
    the  \ktwov, \kthree\ and \ktwor\ features. Top: standard
    profile with two emission peaks and a central depression. Middle:
    profile showing only one emission peak on the red side of the rest
    line-center wavelength. Bottom: profile showing three emission
    peaks. The vertical blue, black and red lines indicate the
    wavelength position of  \ktwov, \kthree\ and \ktwor\ as determined by
    our algorithm.
    \label{fig:prof_labels}}
\end{figure}

Each of the \MgIIhk\ lines, as seen in solar observations, is often
characterized by a central absorption core surrounded by two emission
peaks that are in turn surrounded by local minima. These features are
often refered to as \hthree\ and \kthree\ (line cores), \htwo\ and
\ktwo\ (peaks), and \hone\ and \kone\ (outer minima of peaks). The red
and blue sides of the peaks and minima are distinguished with the
added subscript of R or V respectively. This simplistic description is
at odds with many of the spatially-resolved spectra of our
calculations as the synthetic line profiles can show a more complex
structure. We find line profiles that have from zero to as many as six
emission peaks. Nevertheless, the majority of line profiles shows a
more conventional profile with two outer minima, two emission peaks
and a central minimum. Out of the 254\,016 columns of our 1D
computation, 0.1\% of the profiles did not have any emission peaks,
1.6\% had one peak, 66.6\% had two peaks and 31.7\% had three or more
peaks.

To understand how the spectral properties relate to the physical
properties of the atmosphere, we extracted the
positions of the line cores and red and blue peaks from the
spectra. For each of these features
we extracted the intensity and spectral position (here measured in
Doppler shift relative to the rest wavelength of the line center, for
brevity referred to as `velocity shift'). Throughout the paper we use
the convention that a positive Doppler shift corresponds to a blueshift, and a
positive velocity in the model atmosphere corresponds to an upflow.

Given the large variation of the profile shapes, in some cases some of
these features will not be present, or are difficult to
identify. Also, given the large number of spectra, we needed to use
an automated procedure to extract the quantities. Both the
automated detection and the exotic line profiles introduce
uncertainties in our analysis, of which the reader should be aware.

In the top panel of Figure~\ref{fig:prof_labels} we show a `standard'
profile, with well-defined features. The middle panel shows a profile
with only one emission peak, labeled as \ktwor\ because it has a
wavelength larger than the rest line center wavelength. In this case the
\ktwov\ and \kthree\ features are not defined. The bottom panel shows
a triple-peaked profile. For profiles with three or more peaks it is
not a priori clear which features to select. We detail the assumptions
we used in our extraction algorithm below.

\subsection{Extraction algorithm}

To extract the positions of the line core and peaks we start by using
an extremum-finding algorithm on a small spectral region ($-40 <
\Delta v < 40$\,\kms) around the rest wavelength of each line. This
gives us the wavelengths of all maxima and minima in each line-core
spectrum.

Then we extract the line center position. Depending on
the number of maxima and minima found, we employed a few rules to
obtain an initial estimate of the line center velocity shift. Most of the profiles
have an odd number of minima (usually two maxima and one minimum) --
for these we use the middle minimum. If there are an even number of
minima we use the one with the lowest intensity. If no minima are
found, we use a default value of $\Delta v$=5\,\kms.  Using the estimate
for the line center velocity, a parabolic fit is made to a few
spectral points around the estimate. This yields the line center
velocity and intensity. If these results are too far from the starting
estimate (\ie, an erroneous fit), the spectrum will be marked as outlier (see
below).  To improve the accuracy and purge spurious results, we try to
enforce a smooth spatial distribution of the line centre shifts. Using
a spatial convolution we identify pixels with a significant difference
to the neighboring values. For these `outliers' we redo the line
center fit, using as starting estimate a weighted mean of the surrounding
pixels.

The next part of the procedure is to extract the coordinates of the
red and blue peaks. As for the line center, we use the number of
maxima and minima to estimate the peak locations. Maxima
whose absolute distance to the line center is larger than
30\,\kms\ are discarded. If there are more than four maxima, only the
inner four are used. Whether maxima are assigned as estimates of the red
or blue peaks will depend on their position relative to the line core
(red-ward or blue-ward). If the profile has two
maxima these are used as estimates for the red and blue peaks. If
there are three maxima, we take the closest to the line centre and the
strongest of the remaining. If there are four maxima, we use the inner
maxima unless they are very close together and have low intensities
(\ie, spurious
peaks).  Using the estimates for the peaks, the spectra are
interpolated with a cubic spline on a high resolution wavelength grid. The
velocity shifts and intensities for the peaks are obtained from the
interpolated spectral maxima.


We thoroughly tested the algorithm and in the overwhelming majority of
the cases it found the same spectral features as what a manual visual
inspection would produce. As mentioned before, in certain pixels some
of the spectral features are absent and are marked as such.

\subsection{Overview of results}

\begin{figure*}
  \includegraphics[width=\textwidth]{\figspath/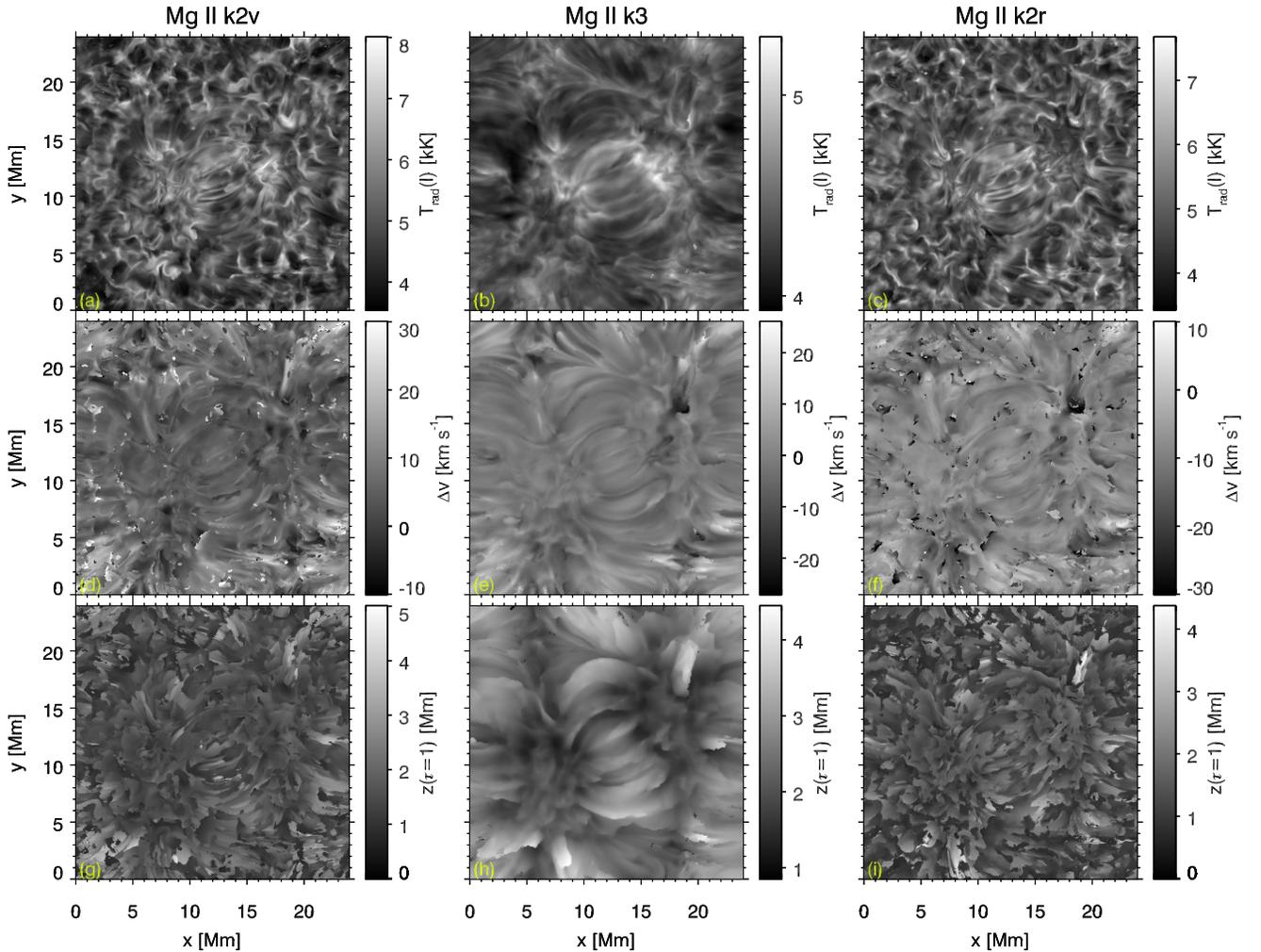}
  \caption{Intensity (top row), Doppler shift (middle row, positive is blueshift) and height
    of optical depth unity (bottom row) in the \ktwov\ (left column),
    \kthree\ (middle column) and \ktwor\ (right column) features. Absent features
    and misidentifications show up as small white and black spots and
    patches, especially visible in panels (d) and (f).
    \label{fig:i_dv_t1}}
\end{figure*}

Figure~\ref{fig:i_dv_t1} shows images for the intensity and Doppler
shift relative to the line-center rest frequency of the features in
the k line, together with the height of optical depth unity at the
feature. The \ktwov\ and \ktwor\ intensity (panels (a) and (c)) show a
pattern reminiscent of chromospheric shocks around the edges of the
image, with a weak imprint of elongated, magnetically formed
structures in the image center. In contrast, the \kthree\ image
(panel (b)) shows elongated structures that are aligned with the
chromospheric magnetic field structure in the whole image.

The second row shows the Doppler shift. Interestingly, the \ktwov\ and
\ktwor\ panels (d) and (f) show elongated large-scale structure, with many
similarities to the \kthree\ Doppler shift (panel (e)). In
Sections~\ref{sec:hk3_diag}--\ref{sec:hk2_diag} we show that the Doppler shift of \ktwov,
\ktwor\ and \kthree\ are all determined by the velocity in the upper
chromosphere, and thus show a correlation.

The bottom row of Figure~\ref{fig:i_dv_t1} shows the height of optical
depth unity. Panels (g) and (i) show that the $\tau\is 1$ height
displays an irregular leaf-like  structure for \ktwov\ and \ktwor. Panel
(h) displays a much smoother structure, again aligned with the
magnetic field lines.

\subsection{On displaying correlations}

\begin{figure*}
  \includegraphics[width=17cm]{\figspath/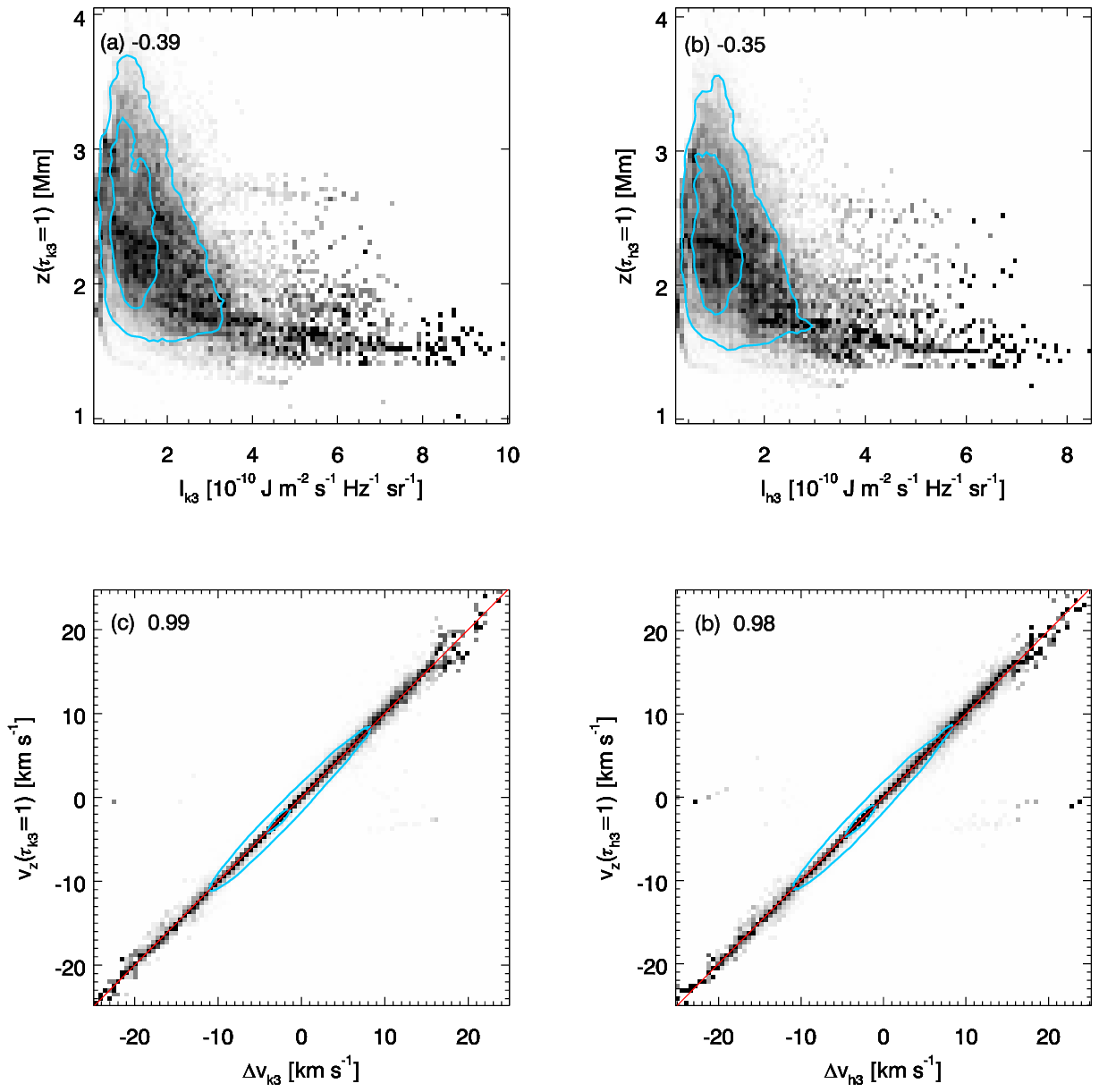}
  \caption{Scaled joint probability density functions (JPDF) of $\tau\is 1$
    height versus the intensity (panels a and b) and vertical velocity at the $\tau\is 1$
    height versus the Doppler shift (panels c and d) of the \kthree\ and
    \hthree\ line centers. The inner blue contour includes 50\% of all
    pixels, the outer contour 90\%.  Each column in the panels is
    scaled to maximum contrast to increase visibility. The Pearson
    correlation coefficient is given in the upper left corner next to
    the panel label. Left column is for the k line, right column for the h line.
 (a)-(b)
    $\tau\is 1$ height of line center versus line center intensity; 
 (c)-(d)
    vertical velocity at the $\tau\is 1$ height of line center (positive is upflow)
    versus the Doppler shift of line center (positive is blueshift), the red line 
    denotes the line $y=x$.
    \label{fig:h3_k3_plot_1}}
\end{figure*}

In Figures~\ref{fig:h3_k3_plot_1}--\ref{fig:h2_k2_plot_3} we show
correlations between synthetic observables and properties of the model
atmosphere. Due to the large number of pixels in our snapshot (63,504
for the 3D computation and 254,016 in the 1D computation), a scatter
plot where each pixel is represented with a black dot would lead to a
completely saturated image. 

Another option is to compute the numerical approximation of the joint
probability density function (JPDF). This approach has the advantage
that saturation does not occur, but the drawback is that the part of
parameter space with a low density is hard to see.

A third option is to compute the JPDF, and then scale each column or
row to maximum contrast, so that each column is a one-dimensional
histogram. This way correlations also show where the JPDF has low
density, but the shape of the original JPDF is lost.

No solution is ideal, so as a compromise we plot the JPDF with each
column scaled to maximum contrast in order to identify correlations
throughout the parameter space, and overplot contours of constant
density in the original JPDF containing 50\% and 90\% of all points to
provide an indication of the shape of the JPDF.

\section{Diagnostic information in h3 and k3} \label{sec:hk3_diag}

In this section we investigate the diagnostic potential of \kthree\ and \hthree\
from their observed quantities: intensities and Doppler shifts. We searched
for correlations between these quantities and the properties of the atmosphere.

\subsection{Line core velocities}

The Doppler shift relative to the rest-frame line center
wavelength contains information on the line-of-sight velocity in
the atmosphere. In panels (c) and (d) of Figure~\ref{fig:h3_k3_plot_1}
we show the correlation between the Doppler shift and the vertical
velocity at optical depth unity. The correlation is tight,
the correlation coefficient is very close to unity,
the FWHM of the $v_z(\tau\is 1)$ velocity distribution at fixed Doppler
shift is typically 0.5\,\kms. This makes \kthree\ and \hthree\ 
excellent velocity diagnostics.

Furthermore, the small difference between the
\kthree\ height of optical depth unity and the height of the
transition region in our model suggests that these spectral features
can be used to measure the velocity in the very upper chromosphere.
This is impossible to do with chromospheric lines observed from the
ground as \Halpha, \CaII\ 854.2\,nm and \CaII\,H\&K, as these lines
have a lower opacity and therefore form further below the transition
region (see Section~\ref{sec:introduction}).

\subsection{Line core intensities}

Panels (b) and (h) of Figure~\ref{fig:i_dv_t1} suggest some
anti-correlation between the intensity and formation height. We
confirm this in panels (a) and (b) of Figure~\ref{fig:h3_k3_plot_1},
which show the JPDF of the
intensity and the height of optical depth unity of \kthree\ and
\hthree. At locations where the intensity is low, the height of optical
depth unity tends to be large. The correlation is weak and the
JPDF displays a large spread. The reason for this anti-correlation is the
3D scattering of the radiation in the chromosphere.

The cores of \MgIIhk\ form in a low density environment with a large
photon mean free path and low photon destruction probability. As a
consequence the local radiation field is smoothed horizontally and, on
average, decreases with height. The Eddington-Barbier relation is
valid, and the emergent intensity is thus approximately equal to the
source function at optical depth unity. The latter is equal to the
angle-averaged radiation field. The source function is low at large $\tau\is 1$
heights, and therefore the emergent intensity is low too.
The same mechanism acts in the \Halpha\ line core, as explained by
\citet{2012ApJ...749..136L}. 
The \MgIIhk\ anticorrelation of formation height and intensity is
weaker than for \Halpha\ mainly for two reasons. First, the \hk\ lines are
resonance lines and lack the mid-chromospheric opacity gap of \Halpha;
they thus have less smoothing of the radiation field. Second, the \hk\
Doppler width is smaller than for \Halpha\ and the source function at
the wavelengths of \kthree\ and \hthree\ is therefore more sensitive to
the velocity field. 

We computed the Pearson correlation coefficient for smaller 2\,Mm $\times$
2\,Mm subfields and found much tighter correlation, with 
correlation coefficients typically between -0.9 and -0.6. This
indicates that at such scales the radiation field is smooth enough to
use the intensity as a relative height measurement. 


 \begin{figure*}
   \includegraphics[width=17cm]{\figspath/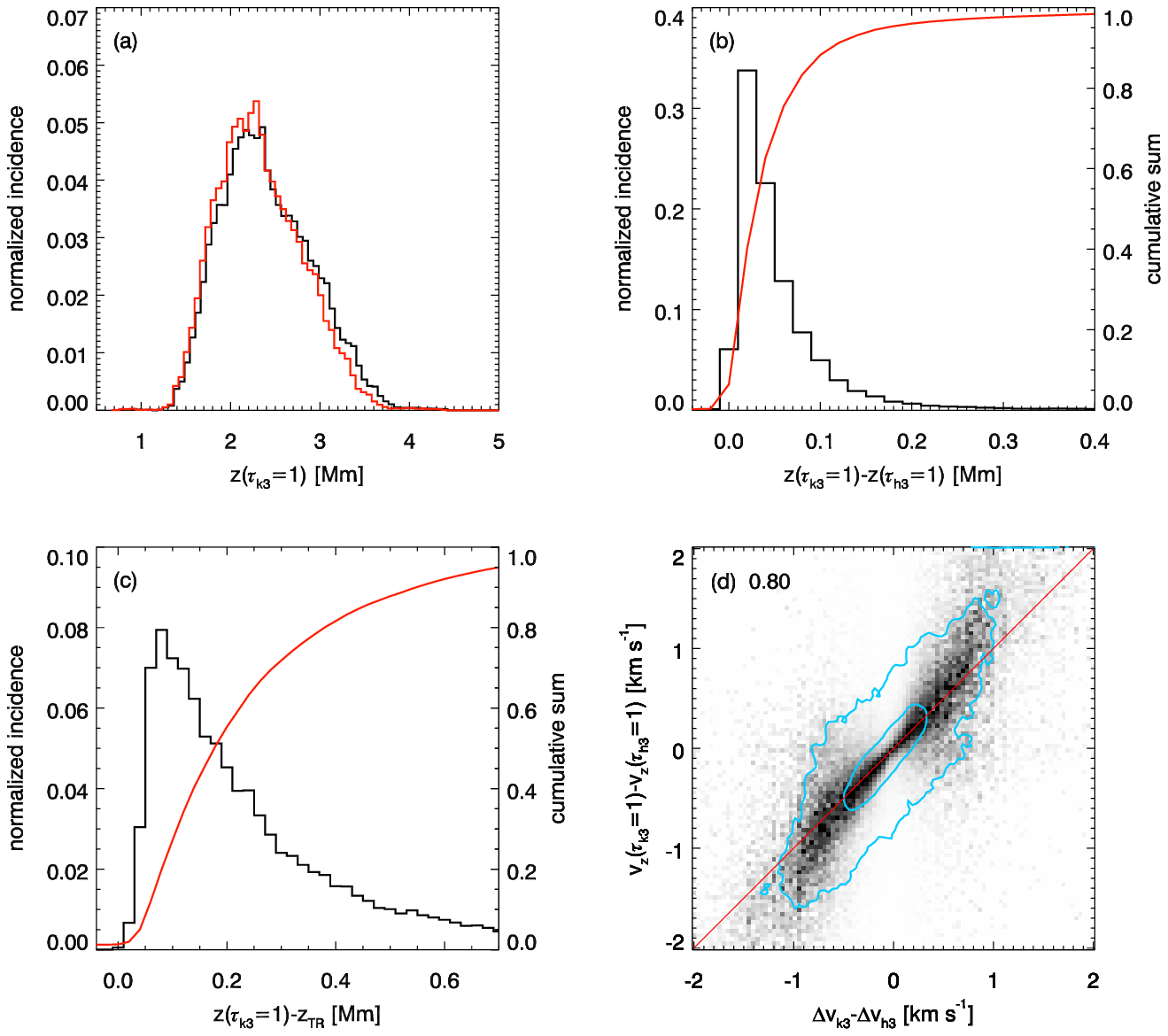}
   \caption{Formation properties of \kthree\ and \hthree.
 (a)
 Histogram of the height of optical depth unity of \kthree\ (black) and
  \hthree\ (red).
 (b) 
  Histogram of difference of the heights of optical depth unity of
     \kthree\ and \hthree\ (black, left-hand scale). The cumulative sum
     of the histogram is shown with the red curve and the right-hand scale.  
 (c)
 Histogram of difference of the height of the transition region in the
 3D model and the height of optical depth unity of
     \kthree\ (black, left-hand scale). The cumulative sum
     of the histogram is shown with the red curve and the right-hand scale;  
(d)
     Scaled joint probability density function of the difference of the
     velocity at optical depth unity of \kthree\ and \hthree\ versus
     the difference in Doppler shift of \kthree\ and \hthree. The
     inner blue contour includes 50\% of all pixels, the outer contour
     90\%.  Each column in the panels is scaled to maximum contrast to
     increase visibility. The Pearson correlation coefficient is given
     in the upper left corner. The red line denotes the line $y=x$.
     \label{fig:h3_k3_plot_2}}
 \end{figure*}

\subsection{Combining \kthree\ and \hthree}

By combining the h and k lines one can extract additional information,
namely the difference in
velocity in the atmosphere between the slightly different heights of
optical depth unity.  In panel (a) of Figure~\ref{fig:h3_k3_plot_2} we
show the histograms of the $\tau\is 1$ height of \kthree\ and \hthree.
The \kthree\ distribution is shifted to slightly larger heights. In
panel (b) this is shown in more detail, where we plot the
histogram of the difference in height of optical depth unity between
\kthree\ and \hthree. The \kthree\ minimum typically forms a few tens
of kilometers higher.

In panel (c) we show where in the chromosphere optical depth unity of
\kthree\ is reached. It shows the histogram of the height difference
between the transition region and $z(\tau_\mathrm{k3}\is 1)$. We
define the height of the transition region in a column in the
atmosphere as  the largest height 
where the temperature drops below 30\,kK. The \kthree\ minimum forms
typically less than 200\,km below the transition region. Finally, in
panel (d) we show the JPDF of the difference in observed Doppler shift
of the h and k minima $\Delta v_\mathrm{k3}-\Delta v_\mathrm{h3}$ and
the difference in velocity at optical depth unity
$v(\tau_\mathrm{k3}\is 1)-v(\tau_\mathrm{h3}\is 1)$. There is a clear
correlation between the two, with a correlation coefficient of 0.80.
The correlation is good enough to detect
the sign and possibly the magnitude of the velocity difference. This opens
up the possibility to detect short-wavelength velocity oscillations
and measure the vertical acceleration of upper-chromospheric material
with an instrument with sufficiently high spatial and spectral resolution.

\section{Diagnostic information of h2 and k2} \label{sec:hk2_diag}

\begin{figure*}
  \includegraphics[width=17cm]{\figspath/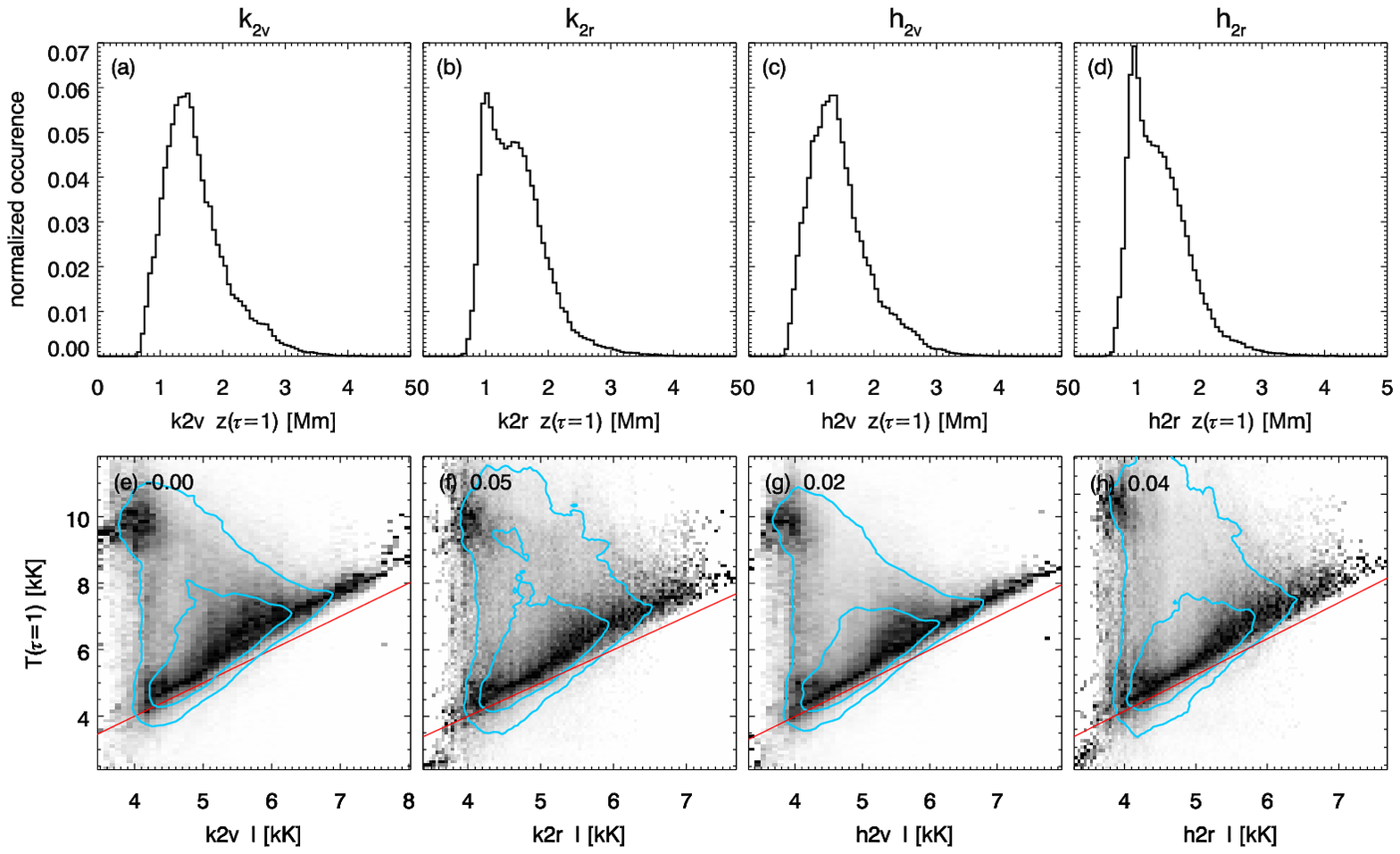}
  \caption{Formation properties of the \MgIIhk\ line-core emission peaks
    Top row (panels (a)--(d)): histogram of the height of optical depth
    unity for the spectral feature indicated above the panels.
    Bottom row (panels (e)--(h)): Scaled joint probability density
    functions the temperature at the height of optical depth unity
    versus the intensity of the emission peaks expressed as a
    radiation temperature.  The inner blue contour includes 50\% of
    all pixels, the outer contour 90\%.  Each column in the panels is
    scaled to maximum contrast to increase visibility. The red lines
    denote the line $y=x$. The Pearson correlation coefficient is
    given in the upper left corner of each panel.
    \label{fig:h2_k2_plot_1}}
\end{figure*}

%
We now turn our attention to the diagnostic information contained in
the intensities and Doppler shifts of the \ktwov, \ktwor, \htwov\ and
\htwor\ peaks. 

\subsection{Peak intensities}

In Figure~\ref{fig:h2_k2_plot_1} we show histograms of the $\tau\is 1$
heights for the  \ktwov, \ktwor, \htwov\ and \htwor\ peaks, 
and JPDFs of the peak intensity and the temperature at their $\tau\is 1$ height.
The distribution of $\tau\is 1$ heights peaks between 1\ and 2\,Mm, with a
tail up to beyond 3\,Mm. This tail is caused by the large variations
of the opacity from velocity variations along the line-of-sight.
The contribution function at the wavelength of the intensity
peaks can therefore be bimodal, with a contribution from the middle
chromosphere and a contribution from higher up in the
chromosphere. The $\tau\is 1$ height can be in either location (see also
Section~\ref{sec:indprof} and Figure~\ref{fig:iform_072_369}). 
The median height of optical depth unity is highest for \ktwov\
(1.46\,Mm), and lowest for \htwor\ (1.29\,Mm). The blue peaks form on
average 60\,km higher than the red peaks.



Panels (e)--(f) show JPDFs of $T(\tau\is 1)$ versus the intensity. The
distributions have a typical structure. For many of the points there 
is a very good correlation between intensity and temperature. For these points
the gas temperature is about 500\,K larger than the radiation temperature.
These are the columns whose $\tau\is 1$ heights are located in the
mid-chromosphere, around 1.5\,Mm. Here the source function is only
partially decoupled from the local temperature, causing the correlation.
This correlation is more tightly constrained for larger intensities.
At lower intensities the distribution widens, with a fraction of the pixels having
a gas temperature higher than the radiation temperature. In these columns
the $\tau\is 1$ heights are located mostly in the upper chromosphere,
where the source function is completely decoupled from the
temperature. Note that the intensity-temperature correlation is
stronger for the blue peaks than for the red peaks.

The correlation shows that large peak intensities are a good
diagnostic of the temperature at the heigth of optical depth
unity. The same holds for a fraction of the pixels with lower peak
intensities, but at low peak intensities there is also a fraction for
which it is not. 


\begin{figure}
  \includegraphics[width=8.8cm]{\figspath/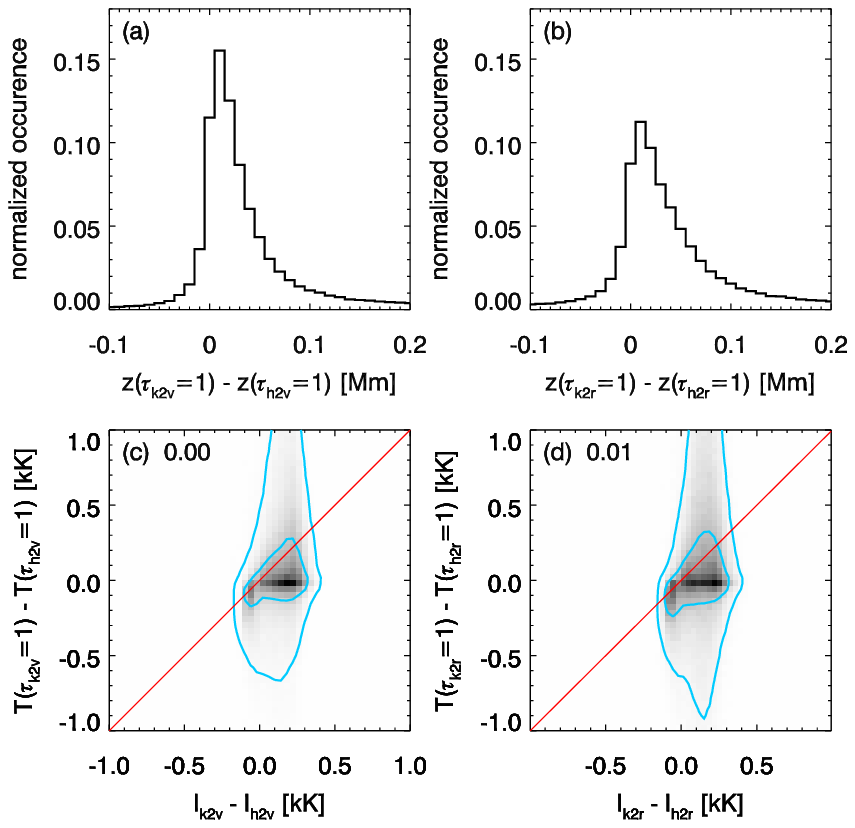}
  \caption{Formation height differences between the emission peaks and
    the correlation between peak intensity and temperature at optical
    depth unity.
(a) histogram of the difference of $\tau\is 1$ height
    of \ktwov\ and \htwov;
(b) same as (a) but for the red peaks;
(c)
    Joint probability density function of the difference of the
    temperature at the $\tau\is 1$ height of \ktwov\ and \htwov\ versus
    the difference of the intensity of \ktwov and \htwov. The inner
    blue contour includes 50\% of all pixels, the outer contour
    90\%. The red line denotes $y=x$. The Pearson correlation
    coefficient is given in the upper left corner of the panel;
(d) same as (c) but for the red peaks.
    \label{fig:h2_k2_plot_2}}
\end{figure}

We also investigated whether it is possible to measure the sign of the
temperature gradient by exploiting the slight difference in $\tau\is
1$ heights between the h and the k line.
We show the results in Figure~\ref{fig:h2_k2_plot_2}. Panel (a) shows
a histogram of the difference in $\tau\is 1$ heights between \ktwov\ and
\htwov. Panel (b) shows the same for the red peaks. In the majority of
cases the peaks of the k line form between 0 and 100\,km higher. 
Panel (c) shows the JPDF of the intensity difference between
\ktwov\ and \htwov, and the difference in temperature at the height of
optical depth unity. There is no usable correlation. The same holds
for the red peaks, as shown in panel (d). The scattering contribution
to the source function adds so much variation that the temperature
differences between the formation heights are washed out in the
emergent intensity.

\begin{figure}
  \includegraphics[width=8.8cm]{\figspath/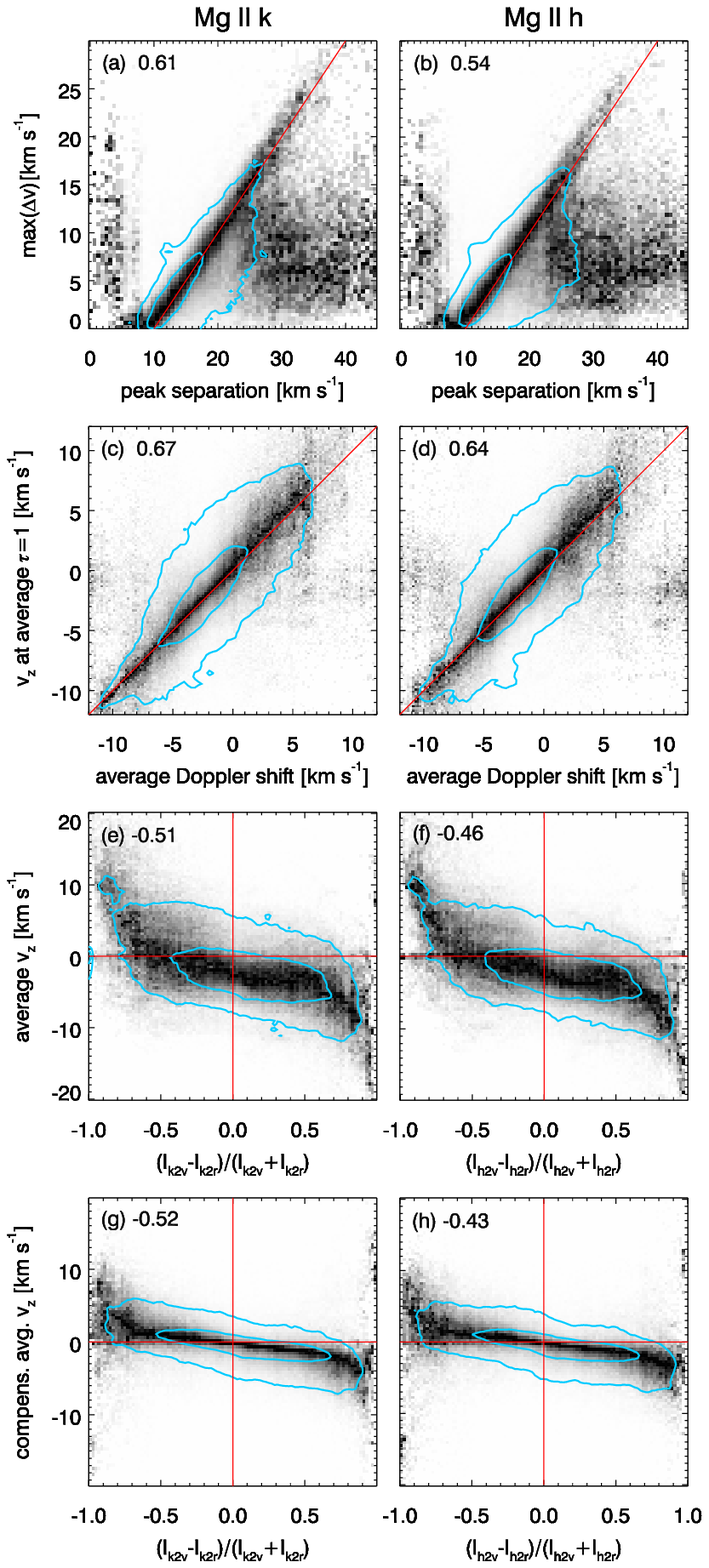}
  \caption{Velocity information contained in \ktwo\ (left column) and 
    \htwo (right column). The correlations are shown as scaled JPDFs. The inner
    blue contour includes 50\% of all pixels, the outer contour
    90\%. Each column in the panels is scaled to maximum contrast to
    increase visibility. The Pearson correlation
    coefficient is given in the upper left corner of each panel.
 (a)-(b)
    velocity difference between the blue and red peaks vs. $\max(\Delta v)$ (see text), 
    the red line is $y=x-10$;
(c)-(d)
    Mean Doppler shift of the blue and red peak vs. the
    vertical velocity at the mean $\tau\is 1$ height of the peaks;
(e)-(h)
    normalized difference of the
    peak intensities vs. the mean vertical velocity
    in the atmosphere between the mean $\tau\is 1$ height of
    the emission peaks and the line core, the red lines denote $x=0$ and $y=0$; 
(g)-(h) same as (e)-(h), but now for the compensated mean vertical velocity (see text).
    \label{fig:h2_k2_plot_3}}
\end{figure}

\subsection{Peak velocities}

The \ktwo\ and \htwo\ peaks form deeper than the line cores, and
thus provide velocity diagnostics for a different height in the
chromosphere. To investigate what information about the atmospheric
velocity is contained in the peaks, we investigated the following observable
quantities: peak separation, average Doppler shifts, and peak intensity
ratios. Figure~\ref{fig:h2_k2_plot_3} shows the results.

The peak separation is a measure of the width of the extinction profile,
and is sensitive to the velocity variations in the upper chromosphere.
At typical chromospheric temperatures the thermal width of the
extinction profile is relatively small, on the order of
2.5\,km\,s$^{-1}$. In an atmosphere without velocity fields this leads to a small peak
separation. Velocity gradients in the atmosphere work to widen the peak separation,
by causing height-dependent wavelength shifts of the narrow thermal extinction profile.
The height-integrated extinction coefficient will then be wider than the
extinction profile at any given fixed height.

For each pixel in the synthetic spectra we measured the peak
separation between the blue and red peaks in each line. Then we
computed $\max(\Delta v)$, the difference between the maximum and
minimum atmospheric velocities in the formation region of the
inter-peak profile (defined as the range between the average $\tau\is
1$ height of the emission peaks and the $\tau\is 1$ height of the line
core minimum). 
%
Panels (a)-(b) show the JPDF of the peak separation and the maximum
velocity difference. There is a clear positive
correlation between the two quantities: a larger peak separation
corresponds to a larger velocity difference in the chromosphere above
the formation height of the peaks. 

A small quantity of points (less than 10\%) does not follow the
near-linear relation between peak separation and $\max(\Delta
v)$. They have a large peak separation and small $\max(\Delta
v)$. These points have two origins. Some points result from a
misidentification from our algorithm.  This typically happens when one
of the peaks is almost non-existent and a smaller spectral feature is
mistakenly assumed to be the peak, or when there are multiple clear peaks and
our algorithm selects the wrong peak.  For the points that have
$\max(\Delta v)$ less than the peak separation minus 15\,\kms, about
half of them result from these misidentifications. The other half is
caused by a naturally wider extinction profile that happens when there
is a temperature maximum in the lower chromosphere
($z\approx0.6-0.9$~Mm). In these cases the source function has a maximum at
heights lower than usual. This results in a larger peak separation due to
the atmospheric temperature structure instead of velocities.

We expect the velocity around the formation heights of the peaks to
influence the wavelength location of the peaks. To investigate this
we computed the  average Doppler shift of the
\ktwov\ and \ktwor\ peaks ($\Delta v_\mathrm{Dop,k}$) as follows:
\be
\Delta v_\mathrm{Dop,k} = -\frac{1}{2} \frac{c}{\lambda_\mathrm{k}}
\left[ (\lambda_\mathrm{k2v}-\lambda_\mathrm{k})+(\lambda_\mathrm{k2r}-\lambda_\mathrm{k})  \right],
\ee
with $c$, $\lambda_\mathrm{k}$, $\lambda_\mathrm{k2v}$ and
$\lambda_\mathrm{k2r}$ respectively the speed of light, the rest-frame
line center wavelength, the observed wavelength of \ktwov\ and the
observed wavelength of \ktwor. With this definition a positive average
Doppler shift corresponds to a blue shift. We then computed the
atmospheric velocity at the average \mbox{$\tau\is 1$} height of the peaks. The
atmospheric velocity is defined such that upflows are positive. In
panel (c) of Figure~\ref{fig:h2_k2_plot_3} we show the JPDF of these two quantities.
There is a clear correlation between the atmospheric velocity and the
Doppler shift, but the distribution shows some spread. Panel
(d) shows the same for \MgIIh.

From 
\citet{1997ApJ...481..500C} 
we know that the \CaII\ \Htwov\ and \Ktwov\ bright grains are
strongest when the atmosphere above the formation height of the
\Htwov\ and \Ktwov\ peaks is moving down. The \MgIIhk\ lines behave
similarly, but exhibit stronger emission peaks because of their shorter
wavelength and higher opacity. As line formation is
almost symmetrical with respect to the line center wavelength, we
expect that an upflow above the emission-peak formation height will
lead to red peaks being stronger than blue peaks.
The ratio of the blue and red peak intensity
might thus be exploited to measure the average velocity in the upper
chromosphere.  It turns out that this is indeed the case.
We computed the intensity ratio of the blue and the red peak as
\be
R_\mathrm{k}=\frac{I_\mathrm{k2v}-I_\mathrm{k2r}}{I_\mathrm{k2v}+I_\mathrm{k2r}},
\ee
so that $I_\mathrm{k2v}>I_\mathrm{k2r}$ yields a positive peak ratio. 
We also computed the average vertical velocity in the atmosphere
between the average $\tau\is 1$ height of the peaks and the $\tau\is 1$
height of \kthree, \ie
\be
  v_\mathrm{avg,k} = \frac{\int_{z1}^{z2} v_z(z') \dd z'}{z_2-z_1}, 
\ee
with 
\be
 z_1=\frac{z(\tau_\mathrm{k2v}\is 1)+z(\tau_\mathrm{k2r}\is 1)}{2}
\ee
 and
\be
 z_2=z(\tau_\mathrm{k3}\is 1). 
\ee
As before, a positive velocity corresponds to upflow. The same procedure 
was followed for the h line.
Panels (e)-(f) of Figure~\ref{fig:h2_k2_plot_3} show the JPDF of
$R_\mathrm{k/h}$ and $v_\mathrm{avg,k/h}$. There is a
correlation. A stronger blue peak corresponds to downflowing material
above the peak formation height, and likewise, a stronger red peak
corresponds to upflow. This is especially clear for large intensity
ratios ($R>0.7$ and $R<-0.7$). For smaller ratios the correlation is weaker.
Note that the majority of the distribution is below
the zero line, indicating that there is on average a downward motion. This
is partly due to a global oscillation in the simulation box (which at the time of this 
snapshot is directed downwards) and partly due to a correlation between density
and velocity: upward moving waves have higher density in the upward phase than in the 
downward phase such that zero average mass-flow gives an average downward velocity.


The spread of the peak-intensity ratio versus velocity correlation can
be significantly reduced by compensating for the velocity at the peak
formation height. This is to be expected, because adding a constant
velocity to the whole atmosphere would shift the line, but not change
the intensity ratio. In panels (g)-(h) we show the peak ratio--velocity
correlation again, but now we subtract the velocity at the average
$\tau\is 1$ height of the peaks from $v_\mathrm{avg,k/h}$. This has two
effects: one, the distribution is shifted upward and passes through the
origin; two, the distribution at a given peak ratio is much
narrower. 

\subsection{Summary}

We summarize our analysis of the intensity and Doppler shift of the
emission peaks:
\begin{itemize}
\item The emission peaks have optical depth unity at around \mbox{$z\is 1.4$\,Mm}, but
with a wide spread. The blue peaks form higher on average than the red
peaks, and the k peaks higher than the h peaks. A fraction of the
points has optical depth unity at larger heights (in the upper
chromosphere).
\item The peak intensity correlates very well with the gas temperature at
optical depth unity, for peak radiation temperatures higher than 6\,kK.
At lower radiation temperatures this correlation has a larger spread,
because more points reach an optical depth unity in the upper
chromosphere, where the source function is completely decoupled from
the  local temperature.
%
\item One cannot exploit the difference in optical depth unity
between the h and k lines to measure the temperature gradient
between these heights. The variation in intensity caused by the
scattering part of the source function is larger than the sensitivity
to temperature.
\item The peak separation correlates well with the
difference between maximum and minimum velocity in the chromosphere
above  \edt{above the height where the intensity peaks are formed}. The average Doppler shift of the blue
and red peaks exhibits a linear correlation with velocity at the average  peak
formation height. 
\item The peak intensity ratio can
be used to constrain the average atmospheric velocity above the peak
formation height: large peak intensity ratios indicate large flow
velocities. Modest intensity ratios correlate with the difference in
velocity at the peak formation height and the average velocity in the
chromosphere above it. 
\end{itemize}

We stress that all correlations have some spread.

\section{Analysis of individual profiles} \label{sec:indprof}

\begin{figure*}
\centering
  \includegraphics[width=12cm]{\figspath/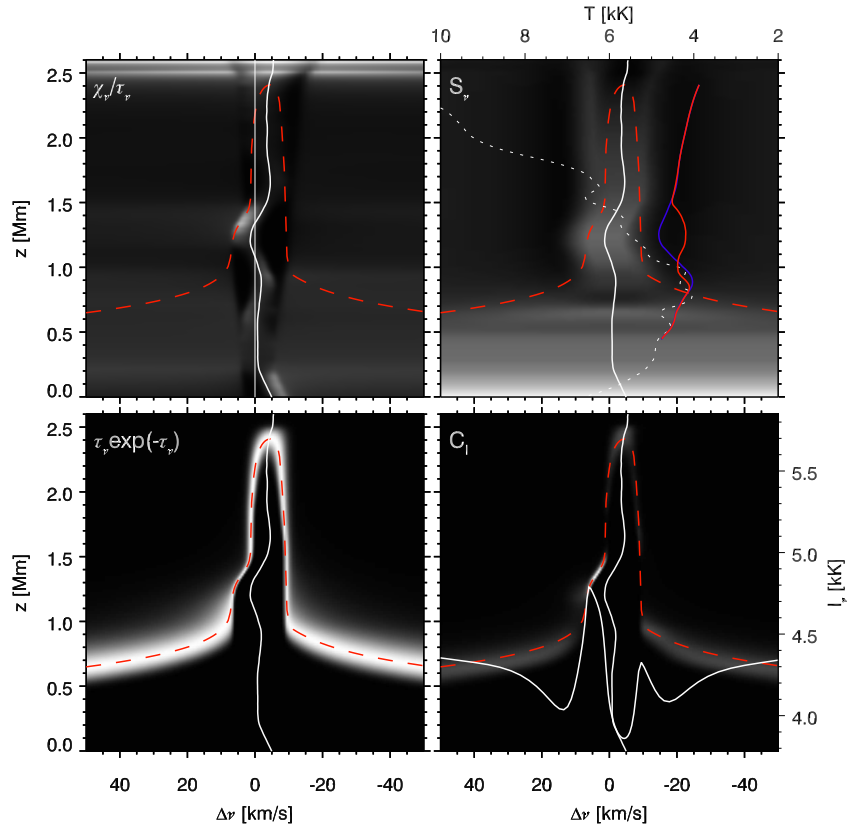}
  \caption{Intensity formation breakdown figure for the
    \MgIIk\ line. Each grey-scale image shows the quantity specified in
    its top-left corner as function of frequency from line center (in
    Doppler shift units) and simulation height $z$.  Multiplication of
    the first three produces the intensity contribution function in
    the fourth panel. A $\tau_\nu\is1$ curve (red dashed) and the
    vertical velocity (white solid, positive is upflow) are
    overplotted in each panel, with a $v_z \is 0$ line in the first
    panel for reference.  The upper-right panel also contains the
    Planck function (dotted) and the line source function along the
    $\tau\is 1$ curve in blue for the part of the $\tau$ curve blueward
    of its maximum value and red for the part on the red side of the
    maximum $\tau\is 1$ height, in temperature units specified along the
    top. The lower-right panel also contains the emergent intensity
    profile, as brightness temperature with the scale along the
    right-hand side. The height scale is cut off where $T=30$\,kK, \ie\
    in the transition region.
    \label{fig:iform_100_100}}
\end{figure*}

\begin{figure*}
\centering
  \includegraphics[width=12cm]{\figspath/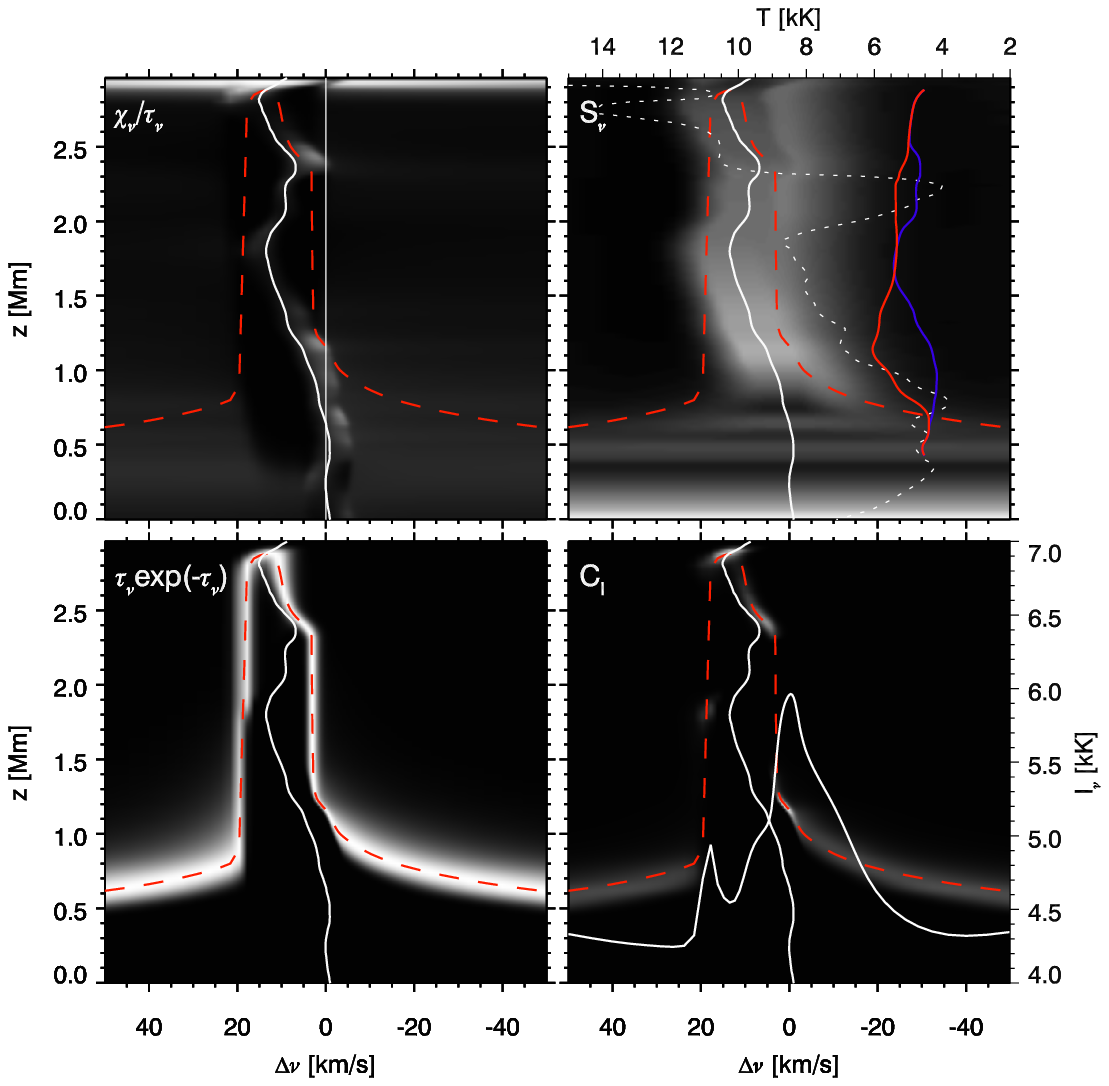}
  \caption{Intensity formation breakdown figure for the
    \MgIIk\ line. The description of the format is given in the caption
    of Fig.~\ref{fig:iform_100_100}
    \label{fig:iform_135_461}}
\end{figure*}

\begin{figure*}
\centering
  \includegraphics[width=12cm]{\figspath/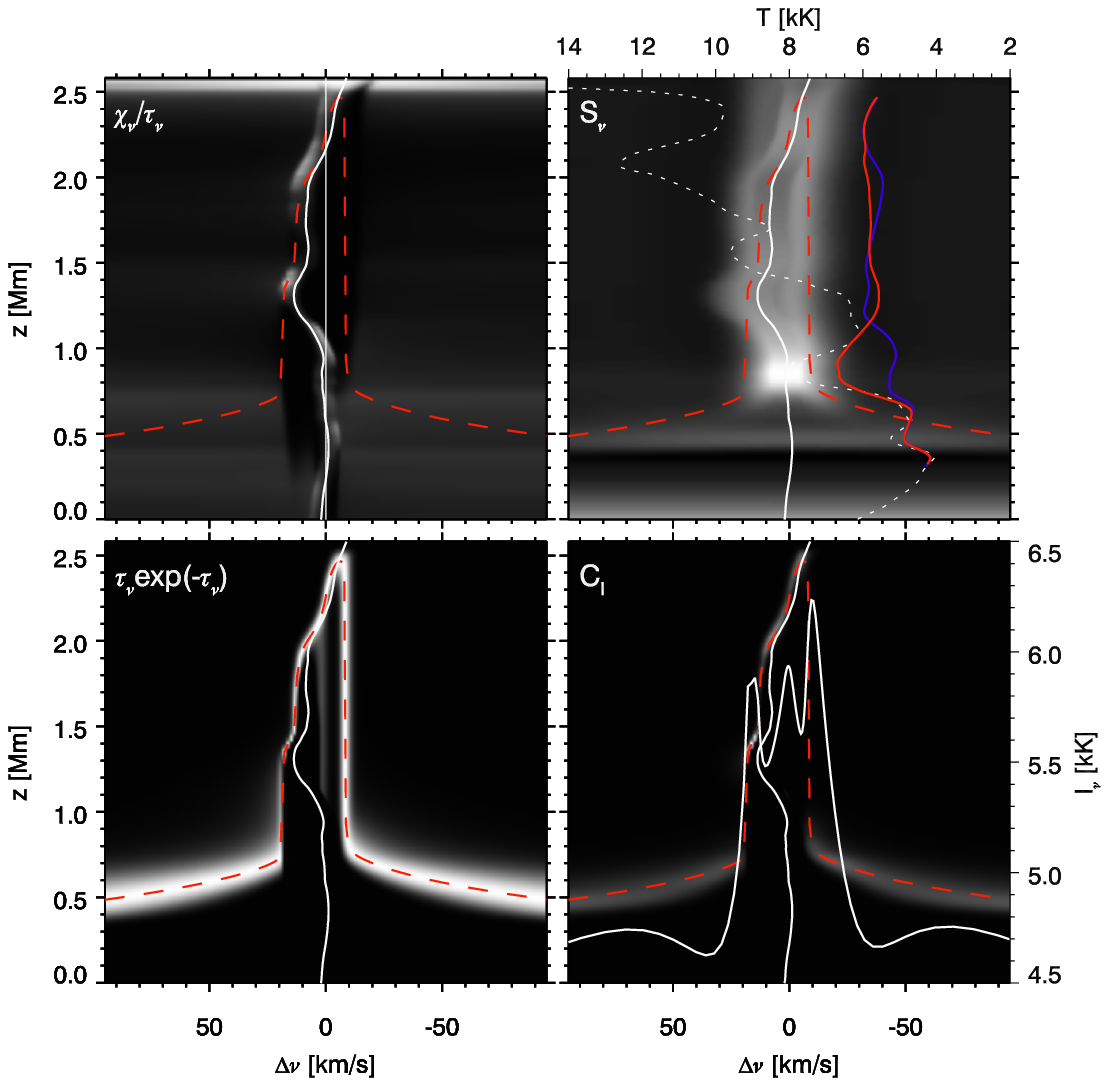}
  \caption{Intensity formation breakdown figure for the
    \MgIIk\ line. The description of the format is given in the caption
    of Fig.~\ref{fig:iform_100_100}
    \label{fig:iform_072_369}}
\end{figure*}

So far we have studied the properties of the emergent \MgIIhk\ line
cores in a statistical manner. In this section we analyze the
formation of the emergent line profile at three different locations in
our 3D atmosphere, and discuss the physical mechanisms behind the
correlations found in Sections~\ref{sec:hk3_diag}
and~\ref{sec:hk2_diag}. These profiles were computed in 1D with \RH.

In Figures~\ref{fig:iform_100_100}--\ref{fig:iform_072_369} we analyze
the formation of the three different \MgIIk\ profiles in detail using
4-panel formation diagrams introduced by Carlsson \& Stein (1994,
1997).
\nocite{1994chdy.conf...47C} 
\nocite{1997ApJ...481..500C} 
In these diagrams the contribution function $C_I$ to the vertically emergent
intensity $I_\nu$ is shown as the product of three terms:
\be
 C_I = \frac{\rmd I_\nu}{\rmd z} = \frac{\chi_\nu}{\tau_\nu} \cdot
S_\nu \cdot \tau_\nu\,\exp^{-\tau_\nu},
\ee
with $z$, $\chi_\nu$, $\tau_\nu$ and $S_\nu$ the height in the
atmosphere, the total extinction coefficient, the optical depth and
the total source function. The latter three are frequency-dependent,
indicated with the $\nu$ subscript. This decompositions shows that the
contribution function to intensity peaks at locations of high opacity
at low optical depth, around optical depth unity and with a high
source function. The figures show these three terms and their product
$C_I$. The large formation height range of the \hk\ lines, the small
thermal line width and the frequency-dependent source function lead to
complex formation behavior. To show the frequency-dependent
source function behavior we add two curves to the source function panel.
They show the source function along the $\tau\is 1$ curve in blue for the 
part of  the $\tau$ curve blue ward of its maximum value and red for the
part on the red side of the maximum $\tau\is 1$ height.

We start with discussing a rather standard line profile, with two
clear emission peaks, the blue peak higher than the red peak, and a
well defined central depression in
Figure~\ref{fig:iform_100_100}. This column in the atmosphere has a
rather weak velocity field, with a slight upflow at 1.25\,Mm and a
downflow of 3--5\,km\,s$^{-1}$ above that height. The temperature has a
minimum at 0.8\,Mm and increases upward, with some additional local minima and
maxima. The largest height of optical depth unity is 2.4\,Mm.
Despite the low velocities, the small Doppler width of the absorption
profile causes well-defined $\chi_\nu/\tau_\nu$ structure, with the latter
showing a prominent peak on the blue side of line-center around
$z\is 1.3$\,Mm. This peak corresponds to the location of the upflow, and
a widening of the $z(\tau_\nu\is 1)$ profile caused by the shift of the
absorption profile at that height towards the blue. Note that on the
red side of line-center the $z(\tau_\nu\is 1)$ height increases very
steeply from 0.9\,Mm to 2.3\,Mm at $\Delta v = -10$\,\kms. 

The source function panel clearly demonstrates the
frequency-dependence of the line source function caused by PRD
effects. The source function decouples from the Planck function at
0.5\,Mm. The blue and red curves are not equal to each other, and show
a complex structure that does not correspond to the temperature
structure, but instead is caused by the interplay of the
temperature, velocity and PRD scattering. The source function on the
blue side of the profile has a prominent maximum at the same location
of the $\chi_\nu/\tau_\nu$ peak caused by the shift of the high
line-core source function towards the blue side of line center. In
contrast, the source function at the same height on the red
$S(\tau\is 1)$ curve shows a minimum because the $\tau\is 1$ curve crosses
the low source function just outside the Doppler core of the
line. Both source function and $\chi_\nu/\tau_\nu$ maxima together
cause the high \ktwov\ peak in the emergent line profile.
This figure illustrates several of the correlations between the
line-core features and the structure of the atmosphere. The
Doppler shift of \kthree\ corresponds exactly to the velocity at the
\kthree\ optical depth unity. The blue \ktwov\ peak is higher than the
red \ktwor\ peak, and indeed the chromosphere shows a downflow between
1.5 and 2.5\,Mm. The red peak is located at the wavelength of the
large jump in $\tau\is 1$ height, in this case the peak $\tau\is 1$ height
is located at $z=0.9$\,Mm, but might as well have been located above
2\,Mm. 

In Figure~\ref{fig:iform_135_461} we show the formation breakdown of a
profile with a higher \ktwor\ than \ktwov\ peak. This column shows
upflows up to 15\,\kms from 0.7\,Mm and upwards. The chromosphere is
rather hot between 1 and 2\,Mm height, and a prominent shock with a
temperature of 14.5\,kK at 2.2\,Mm is present just below the
transition region. The chromosphere is moving up, showing the
correlation between average chromospheric velocity and peak intensity
ratio. The blue intensity maximum has a doubly-peaked contribution
function, one contribution is from $z=1.8$\,Mm and another one from
2.4\,Mm. The red peak originates from 1.2\,Mm height. The Doppler
shift of \kthree\ again is equal to the velocity at optical depth
unity. The average Doppler-shift of the \ktwov\ and \ktwor\ peaks is
9.3\,\kms, and the vertical velocity at the average optical depth
unity of the peaks is 10.5\,\kms.

In Figure~\ref{fig:iform_072_369} we show the formation of a very
complex line-core profile that shows five emission peaks. The two
outermost small peaks at $\Delta \nu = \pm 70$\,\kms\ are caused by a local
maximum in temperature between 0.4 and 0.5\,Mm height. Note that even
at this low height the source function is already partially decoupled
from the local temperature. The highest peak, at $\Delta \nu =
-10$\,\kms, is caused by a local maximum in temperature at 0.9\,Mm
height. The peak at $\Delta \nu = 15$\,\kms\ is caused by a maximum in
the $\chi_\nu/\tau_\nu$ term; the source function along the $\tau\is 1$
curve around the peak formation height is rather flat. Following the
$\tau\is 1$ curve upward the source function goes though a minimum at
1.9\,Mm causing the central depression at $\Delta \nu = 10$\,\kms. The
source function then reaches a maximum at 2.3\,Mm causing the central
emission peak at $\Delta \nu = 0$\,\kms. Even higher up the source
function decreases again, leading to the intensity depression at
$\Delta \nu = -5$\,\kms. Note that the emission peaks at $\Delta \nu =
15$\,\kms and 0\,\kms and the depressions at 10\,\kms\ and -5\,\kms
are not caused by the temperature structure at optical depth
unity. Instead they are caused by the atmospheric velocity causing
modulation of the $\chi_\nu/\tau_\nu$ term and the
frequency-dependence 
of the source function caused by PRD effects. 

The formation of this line profile again illustrates the intricate
formation properties of the \MgIIhk\ line cores. Realistic model
atmospheres with complex velocity and temperature structure can
produce profiles that exhibit more than two emission peaks, with peak
intensities that do not necessarily correspond to increases in the gas
temperature at optical depth unity. The main complicating factor is
PRD, which introduces a frequency-dependence to the source function
that is very difficult to predict without performing the actual
radiative transfer computation.

\section{Comparison with observations}            \label{sec:obs_sim_comp}

\begin{figure}
  \includegraphics[width=8.8cm]{\figspath/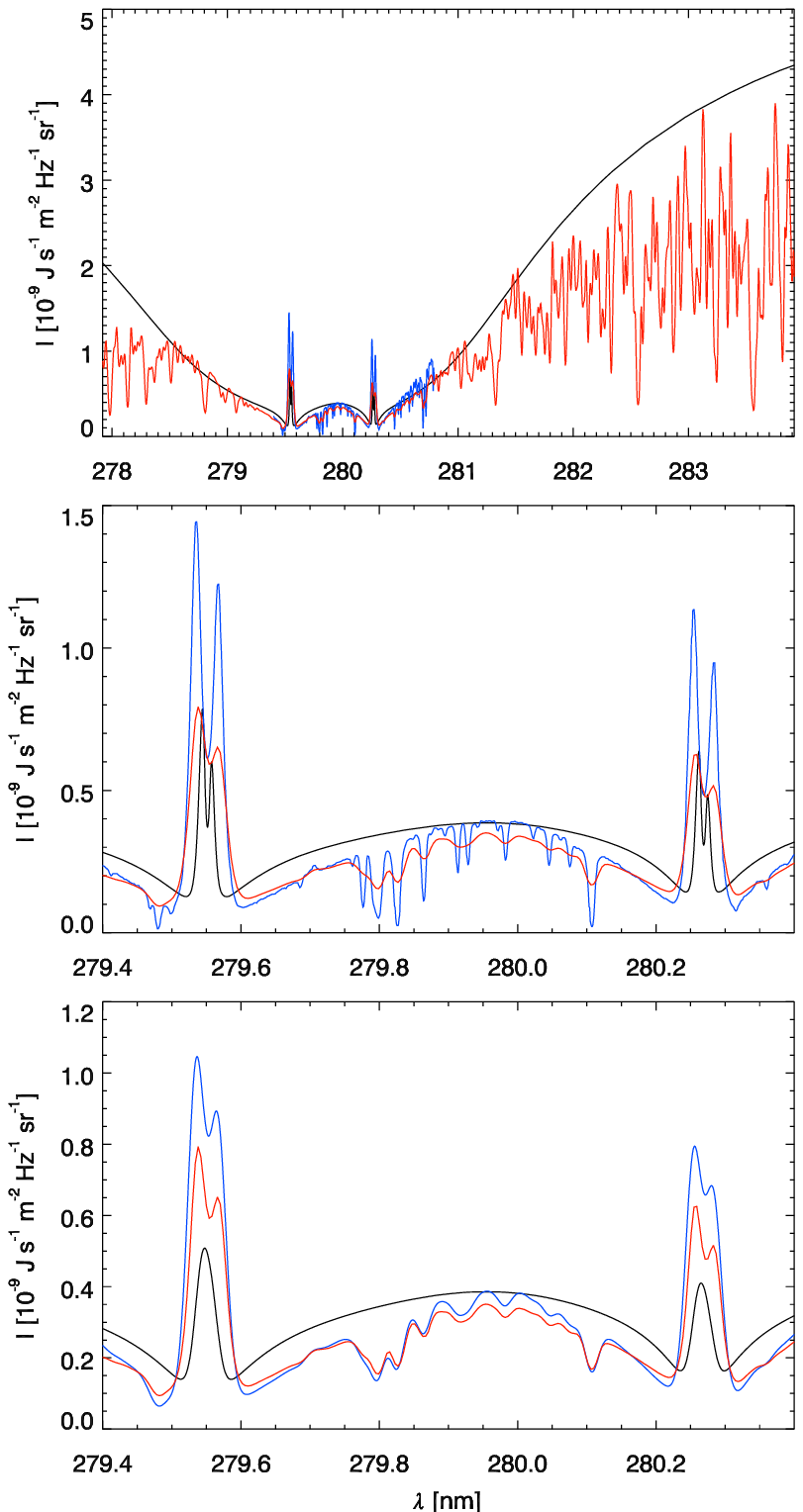}
  \caption{Comparison of the average spectrum from our 1D computation
    with \RH\ (black) with observations of solar disk center with the
    RASOLBA balloon experiment \citep[blue,][]{1995A&A...295..517S} and
    observations of the quiet sun at disk center during the ninth
    flight of the HRTS sounding rocket experiment
    \citep[red,][]{2008ApJ...687..646M}. The top panel has a wavelength
    range including a part of the line wings. The middle and bottom panels show only the
    \hk\ line core region. In the bottom panel the RASOLBA and synthetic spectra were
    convolved to match the spectral resolution of HRTS.
    \label{fig:obs_sim_comp}}
\end{figure}

In order to assess whether the synthetic line profiles resemble the
Sun we compare them with observations. Observations with spatial
resolution comparable to the simulations are not available, so we
chose to compare the spatially averaged profile from the simulation to
high spectral resolution observations.  \edt{We emphasize that the
  goal of this paper is not to reproduce the spatially averaged
  observed profile, but rather to investigate the sensitivity of
  spatially resolved Mg II line profiles to variations in the solar 
  atmosphere.}

In Figure~\ref{fig:obs_sim_comp} we show the spatially averaged
synthetic \MgIIhk\ spectrum from our 1D column-by-column calculation
\edt{(without microturbulence)}
with observations of solar disk center
with the RASOLBA balloon experiment 
\citep{1995A&A...295..517S}
and quiet sun at disk
center with the HRTS9 sounding rocket experiment
\citep{2008ApJ...687..646M}.
The synthetic line profile does not include the blends in the
\hk\ wings, so it is artificially smooth.

\citet{1995A&A...295..517S} specify that the RASOLBA data were
taken at disk center but do not state whether the target region was
quiet sun, network, plage or a combination thereof, whereas the HRTS9
observation targeted quiet sun. Because the central emission peaks and
the wing intensity in RASOLBA are stronger than in 
 HRTS9, we suspect that the RASOLBA observations included
more network or plage than HRTS9.

The synthetic line wings follow the upper envelope of the HRTS9 data
quite well up to 283\,nm (top panel). For larger wavelengths the
synthetic intensity is higher, possibly caused by the neglect of the
broad wings of the \MgI\ line at 285.2\,nm in the synthetic spectrum.
The middle and lower panels of Figure~\ref{fig:obs_sim_comp} show the line core region.
In the lower
panel we smeared the RASOLBA and synthetic spectra to the 20~pm resolution of
HRTS9 to make a better comparison.
The simulation predicts weaker and narrower emission profiles than observed, 
but correctly predicts stronger blue peaks and weaker red peaks.
The smaller peak separation suggests that the simulation
has a weaker velocity field than the Sun. At the resolution of HRTS9, the synthetic 
red and blue peaks blend into a single peak. The lower peak intensity and
integrated emission suggest that the simulation has 
mid-chromospheric temperature lower than the Sun (see Section~\ref{sec:hk2_diag}). 
Both of these effects may also be related to the magnetic field
distribution of the simulation. \edt{In Section~\ref{subsec:lim_mod} we
briefly discuss what is missing in the models.  A more
extensive investigation of what is missing should await the much
higher quality observations that the IRIS mission will provide}.

The weaker and narrower mean emission profile from the simulation
means that our sample of line profiles most likely cover a more
limited range of physical conditions than those that occur in the
Sun. Nevertheless, we believe our sample is still extensive enough to
cover a good variety of conditions present in at least the quiet Sun.

\section{Discussion and conclusions}            \label{sec:conclusions}

The \MgIIhk\ lines show a complex formation behaviour that spans the
entire chromosphere. The variations in temperature, density and
velocity in the chromosphere cause a profuse variety of resulting line shapes.
Based on statistical-equilibrium non-LTE radiative transfer
computations from a snapshot of a 3D radiation-MHD simulation we
sought to identify how the \MgIIhk\ spectra relate to the underlying
atmosphere.

\subsection{Limitations of the model atmosphere} \label{subsec:lim_mod}

We made use of one of the most realistic simulations of the solar chromosphere
currently available and modern non-LTE radiative transfer codes that account
for partial redistribution effects. Nevertheless, our approach is not without
caveats. 

The simulation used to compute our model snapshot has some limitations.
First, it has a limited spatial resolution. Test computations
of simulations with the same domain size but double resolution in each
dimension show that higher resolution simulations yield more violent
dynamics and significantly more small-scale structure. This is not
accounted for in the present paper. Second, we study only one
snapshot with one magnetic field configuration. Different field
configurations may yield different results. Third, the numerical
simulation does not include all physical processes that are important
in the chromosphere. Two processes in particular are ignored that have a
significant effect on the thermal structure in the chromosphere:
non-equilibrium ionization of helium
  \citep{2011A&A...530A.124L}
and the effect of partial ionization on chromospheric heating by
magnetic fields
  \citep{2012ApJ...747...87K,2012ApJ...753..161M}.

A limitation of our radiative transfer calculations is the assumption of 
statistical equilibrium. This assumption is based on the analysis of
the ionization/recombination timescale in Paper~I, where we found that
in the upper chromosphere this timescale is of the order of
50\,s, and much smaller in deeper layers. Our results are therefore
not necessarily valid in circumstances where the thermodynamic state of
the atmosphere changes over shorter timescales, such as in
flares
\citep[\eg][]{2011SSRv..158....5H}
and type II spicules
\citep[\eg][]{2007PASJ...59S.655D, 2012ApJ...759...18P}.
Computations without the assumption of statistical equilibrium are
currently only possible using 1D simulations
\citep{2002ApJ...572..626C,2003ApJ...589..988R,2009A&A...499..923K},
and therefore lack realism in other areas. Nevertheless it would be
interesting to study the response to rapid heating events of the
\MgIIhk\ lines with time-dependent radiative transfer.

\subsection{The \kthree\ and \hthree\ intensity minima}

The most powerful diagnostic from \kthree\ and \hthree\ is their
velocity shift.  We found it to be tightly correlated with the
atmospheric velocity at $z(\tau \is 1)$.  This correlation could be
affected if the temperature in the upper chromosphere is higher in the
Sun than in the simulation. The tight correlation depends on the steep
optical depth-height gradient, that causes a very narrow height
interval with non-zero contribution function. A temperature increase
could cause significant ionization to \MgIII, and in some cases a
corresponding decrease in optical depth-height gradient, and thus a
larger formation height range over which the atmospheric velocity is
averaged. 

The anticorrelation of the \kthree\ and \hthree\ intensities and $\tau
\is 1$ height also offers significant diagnostic potential. The
intensity can be used as a measure of the relative height of optical
depth unity for locations lying a few Mm or less apart. Combined with
the fact that optical depth unity in the intensity minima is typically
located less than 200 km below the transition region, they could
provide a measure of the local shape of the transition region.

This anticorrelation is likely not qualitatively affected by
differences in the upper chromosphere between the Sun and our
simulation, because the average radiation field that sets the
correlation is not affected by the local temperature. Changes in the
temperature around the thermalization depth will change the shape and
location in parameter space of the distribution. However, because the
intensity can only be used as a relative, and not absolute, height
diagnostic this does not affect its usefulness.

\subsection{The \ktwo\ and \htwo\ intensity maxima}


The \ktwo\ and \htwo\ intensities can be used as temperature
indicators. Figure~\ref{fig:h2_k2_plot_1} illustrates that a strong
correlation between peak intensity and temperature at the $\tau \is 1$
height exists for radiation temperatures above 6~kK. For radiation
temperatures below this value the distributions show a significant
amount of scatter, caused by points where the peaks are formed in the
upper chromosphere where the source function is decoupled from the
local temperature.

For pixels with a good correlation the $\tau \is 1$ height cannot be
given because of the spread in $\tau \is 1$ heights and the presence
of multiple-peaked contribution functions (see
Figures~\ref{fig:iform_100_100} and~\ref{fig:iform_135_461}). Still,
we find that a strong emission peak indicates a temperature typically
500 K higher than the peak radiation temperature at some height in the
mid-chromosphere. The time variation of the peak intensity can thus be
used as an indicator of temperature variations. Note that even the
quiet Sun observations show larger average peak intensities than our
simulation. It is therefore reasonable to expect that spatially
resolved observed spectra will typically have more pixels with high
peak intensity for which the intensity-temperature correlation is
good than our simulation.


The spatial average of the peak intensity can be used as an
observable to test models of the chromosphere. The average peak
intensity increases if the temperature in the mid-chromosphere is
higher. Large discrepancies between observed and modeled average peak
intensity therefore indicate an incorrect temperature structure in the
models. They can therefore indicate whether the models include
sufficient chromospheric heating and a correct description of the
equation of state.

We found that the peak separation is mainly determined by the
large-scale velocity gradients in our simulation. 
The correlation between peak
separation and the line-of-sight velocity gradient is good enough that
the peak separation can be used both diagnostically as a measure of
the velocity structure in the upper chromosphere and as a test of
chromospheric models.

The comparison with observations show that the average synthetic peak
separation is smaller than observed. We speculate that in the Sun also
turbulence at lengths too small to be resolved in our simulation might
play a role.  An interesting follow-up study would be to investigate
peak separation in a series of models with identical parameters but
for different spatial resolution, to see how the peak separation
changes.

The average Doppler shift of the peaks gives a rough indication of the
vertical velocity at the average $\tau \is 1$ height of the peaks.

Finally, the intensity peak height ratio is an indication of the
average velocity between the peak $\tau \is 1$ height and the $\tau
\is 1$ height of the intensity minimum. This relation is very clear
for large intensity ratios: very asymmetric peak heights indicate
strong upflows or downflows. This indicates they are suited as a
diagnostic of both internetwork acoustic waves, just as  \CaII\,H\&K\ 
\citep{1997ApJ...481..500C},
and as diagnostics of compressive waves in more strongly magnetized
regions in the solar atmosphere.

\subsection{\MgIIhk\ and IRIS}

So far we studied the \MgIIhk\ lines at the native spatial
resolution of the simulation and high spectral resolution of the
non-LTE radiative transfer computations. IRIS will be equipped with an
imaging spectrograph with $<$8\,pm spectral resolution and a
\MgIIk\ slit-jaw imager with 0.4\,nm\ band pass. Both
instruments will have a spatial resolution of $0.\arcsec4$. 
The finite resolution of IRIS will influence the diagnostic properties
of \MgIIhk\ and the correlations derived in the current paper.
In forthcoming papers of this series we will quantify these effects
and also study the time evolution of synthetic imagery and spectra computed 
from a time series of 3D radiation-MHD snapshots.

\begin{acknowledgements}
    JL recognizes support from the Netherlands Organization for
   Scientific Research (NWO).  This research was supported by the
   Research Council of Norway through the grant ``Solar Atmospheric
   Modelling'' and through grants of computing time from the Programme
   for Supercomputing, by the European Research Council under the European 
   Union's Seventh Framework Programme (FP7/2007-2013) / ERC Grant 
   agreement nr. 291058, and by the computing project s1061 from the High End
   Computing Division of NASA. TMDP was supported by the NASA Postdoctoral
    Program at Ames Research Center (grant NNH06CC03B).
   BDP acknowledges support from NASA grants NNX08AH45G, 
   NNX08BA99G, NNX11AN98G, and NNG09FA40C (IRIS).
\end{acknowledgements}

\end{document}